\documentclass[runningheads]{llncs}
\usepackage{graphicx}
\usepackage{subcaption}
\usepackage{wrapfig}
\usepackage[misc]{ifsym}

\begin{document}
%

%\title{Hybrid Collaborative Filtering Ontology-Based Model for Recommending Chemical Compounds \thanks{Supported by organization Funda\c{c}\~ao para a Ci\^encia e a Tecnologia (FCT), under LASIGE Strategic Project - UID/CEC/00408/2019, CENTRA Strategic Project UID/FIS/00099/2013, FCT funded project PTDC/CCI-BIO/28685/2017 and PhD Scholarship SFRH/BD/128840/2017.}}

\title{Hybrid Semantic Recommender System for Chemical Compounds\thanks{This work was supported by the Funda\c{c}\~ao para a Ci\^encia e Tecnologia (FCT), under LASIGE Strategic Project - UID/CEC/00408/2019, UIDB/00408/2020, CENTRA Strategic Project UID/FIS/00099/2019, FCT funded project PTDC/CCI-BIO/28685/2017 and PhD Scholarship SFRH/BD/128840/2017.}}
\titlerunning{Hybrid Semantic Recommender System for Chemical Compounds}
% If the paper title is too long for the running head, you can set
% an abbreviated paper title here

%%%%%%%%%%%%%%%%%%%
\author{M\'arcia Barros \Letter\inst{1,2}\orcidID{0000-0002-9728-9618}
\and Andr\'e Moitinho\inst{2}\orcidID{0000-0003-0822-5995} 
\and Francisco M. Couto\inst{1}\orcidID{0000-0003-0627-1496}}

\authorrunning{M. Barros et al.}

% First names are abbreviated in the running head.
% If there are more than two authors, 'et al.' is used.
%
\institute{LASIGE, Departamento de Inform\'atica, Faculdade de Ci\^encias, Universidade de Lisboa, 1749--016 Lisboa, Portugal. \email{mcbarros@fc.ul.pt} \and
CENTRA, Departamento de F\'isica, Faculdade de Ci\^encias, Universidade de Lisboa, 1749--016 Lisboa, Portugal\\}

\maketitle              % typeset the header of the contribution
\begin{abstract}
Recommending Chemical Compounds of interest to a particular researcher is a poorly explored field. The few existent datasets with information about the preferences of the researchers use implicit feedback. The lack of Recommender Systems in this particular field presents a challenge for the development of new recommendations models. In this work, we propose a Hybrid recommender model for recommending Chemical Compounds. The model integrates collaborative-filtering algorithms for implicit feedback (Alternating Least Squares (ALS) and Bayesian Personalized Ranking(BPR)) and semantic similarity between the Chemical Compounds in the ChEBI ontology (ONTO). We evaluated the model in an implicit dataset of Chemical Compounds, CheRM. The Hybrid model was able to improve the results of state-of-the-art collaborative-filtering algorithms, especially for Mean Reciprocal Rank, with an increase of 6.7\% when comparing the collaborative-filtering ALS and the Hybrid ALS\_ONTO.

\keywords{Recommender System  \and Implicit feedback \and Ontology \and Collaborative-Filtering \and Semantic similarity.}
\end{abstract}
\section{Introduction}
\label{Sec:intro}
The recommendation of Chemical Compounds of interest for scientific researchers has not been widely explored~\cite{ishihara2015identification,seko2018compositional}. However, Recommender Systems (RSs) may help in the discovery of compounds, for example, by suggesting items not yet studied by the researchers. One challenge in this field is the lack of available datasets with the preferences of the researchers about the Chemical Compounds for testing the RS. More recently, alternatives have emerged with the development of datasets consisting of data collected from implicit feedback. Unlike what happens with other datasets, for example, Movielens~\cite{harper2015movielens}, these datasets do not contain the specific interests of the researchers. Instead, this information is extracted from the activities of the researchers, for example, through scientific literature~\cite{ortega2018artificial,barros2019using}.  

Datasets of explicit or implicit feedback require different recommender algorithms, especially because implicit feedback has some significant downgrades, such as the lack of negative feedback, and unbalanced ratio of positive vs unobserved ratings~\cite{rendle2009bpr,khawar2019conformative}. When dealing with implicit feedback datasets, the solution involves applying learning to rank (LtR) approaches. LtR consists in, given a set of items, identify in which order they should be recommended~\cite{rendle2009learning}. 

The main approaches in RSs are Collaborative-Filtering (CF) and Content-Based (CB)~\cite{ricci2015recommender}. CF uses the similarity between the ratings of the users, and CB uses the similarity between the features of the items. CF approaches cannot deal with new items or new users in the system, i.e., items and users without ratings (cold start problem). CB does not need to deal with this problem for new items, and that is the main reason Hybrid RSs (CF + CB) exist. One of the tools used by CB are ontologies~\cite{tarus2018knowledge}, which are related vocabularies of terms and definitions for a specific field of study \cite{uschold1996ontologies,barros2016knowledge}. Some examples of well-known ontologies are the Chemical Entities of Biological Interest (ChEBI)\footnote{\url{https://www.ebi.ac.uk/chebi/}}~\cite{hastings2015chebi}, the Gene Ontology (GO)\footnote{\url{http://geneontology.org/}}~\cite{gene2018gene}, and the Disease Ontology (DO)\footnote{\url{http://disease-ontology.org/}}~\cite{schriml2018human}.

In this paper, we propose a Hybrid recommender model for recommending Chemical Compounds, consisting of a CF module and a CB module. In the CF module we tested two algorithms for implicit feedback datasets, Alternating Least Squares (ALS)~\cite{hu2008collaborative} and Bayesian Personalized Ranking (BPR)~\cite{rendle2009bpr}, separately. In the CB module we explored the semantic similarity between the compounds in the ChEBI ontology (ONTO algorithm). The Hybrid model combines ALS + ONTO, and BPR + ONTO. The framework developed for this work is available at \url{https://github.com/lasigeBioTM/ChemRecSys}.

\section{Related Work}
\label{Sec:rw}
There are a few studies using RS for recommending Chemical Compounds. \cite{ishihara2015identification} describes the use of CF methods for creating a Free-Wilson-like fragment recommender system. \cite{seko2018compositional} use RS techniques for the discovery of new inorganic compounds, by applying machine-learning to find the similarity between the proposed and the existent compounds. 

Next, we describe studies using ontologies for improving the performance of CF algorithms. 
\cite{liao2010library} created a RS for recommending English collections of books in a library. The authors developed PORE, a personal ontology Recommender System, which consists of a personal ontology for each user and then the application of a CF method. They used a standard normalized cosine similarity for finding the similarity between the users. 
\cite{sieg2010improving} also used an ontology for creating users' profiles for the domain of books. They calculated the similarity, not between the ratings of the users, but based on the interest scores derived from the ontology. The CF method used was the k-nearest neighbours. 
\cite{shambour2012trust} developed a Trust–Semantic Fusion approach, tested on movies and Yahoo! datasets. Their approach incorporates semantic knowledge to the items primary information, using knowledge from the ontologies. They used the user-based Constrained Pearson Correlation and the user-based Jaccard similarity. 

\cite{ostuni2013top} presented a solution for the top@k recommendations specifically for implicit feedback data. The authors developed the Spank - semantic path-based ranking. They extracted path-based features of the items from DBpedia and used LtR algorithms to get the rank of the most relevant items. They tested the method on music and movies domains. 
\cite{al2015semantic} developed a new semantic similarity measure, the Inferential Ontology-based Semantic Similarity. The new measure improved the results of a user-based CF approach, using Pearson Correlation for calculating the similarity between the users. The authors tested the approach on the tourism domain.
Most recently, \cite{nilashi2018recommender} developed a Hybrid RS tested on the movies domain. The method used Single Value Decomposition for dimensionality reduction for the item and user-based CF, and ontologies for item-based semantic similarity, improving the CF results. They do not deal with implicit data.

To the best of our knowledge, our study is the first to use semantic similarity for recommending Chemical Compounds, dealing with implicit data by using state-of-the-art methods (ALS and BPR) and improving the results for the top@k in several evaluation metrics.  

\section{The Proposed Model}
\label{Sec:meth}
\begin{figure}[ht]
\includegraphics[width=1\textwidth]{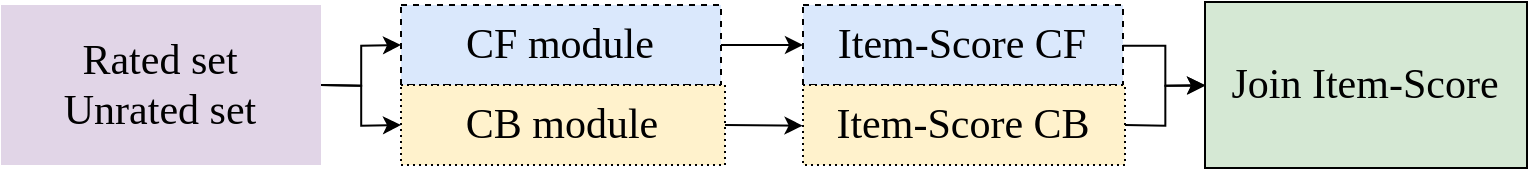}
\caption{Workflow of the Hybrid recommender model.}
\label{fig:flow}
\end{figure}

The proposed model has two modules: CF and CB. Figure \ref{fig:flow} shows the general workflow of the model. 
The input data used in this model has the format of \textless user,item,rating\textgreater. The unrated set represents the items we want to rank to provide the best recommendations in the first positions to a user. The rated set are the items the users already rated. Since we will split the data into train and test, lets call train set to the rated set and test set to the unrated set. Both train and test sets are the input for CF and CB modules.  Using CF algorithms for implicit feedback datasets, the CF module gives a score for each item in the test set. The CB module uses semantic similarity for providing a score for the items in the test set. In the last step, the scores from CF and CB modules are combined and sorted in descending order. 

For the CF module, we selected state-of-the-art CF recommender algorithms for implicit data \footnote{\url{https://implicit.readthedocs.io/en/latest/index.html}}, ALS~\cite{hu2008collaborative} and BPR~\cite{rendle2009bpr}. ALS is a latent factor algorithm that addresses the confidence of a user-item pair rating. BPR is also a latent factor algorithm, but it is more appropriate for ranking a list of items. BPR does not just consider the unobserved user-item pairs as zeros, but instead, it takes into consideration the preference of a user between an observed and an unobserved rating. 
%CF module receives as input the test items for a given user and the data for training the model with ALS or BPR algorithms. The output is a score (S) for each item.

\begin{wrapfigure}{R}{0.4\textwidth}
\centering
\includegraphics[width=0.4\textwidth]{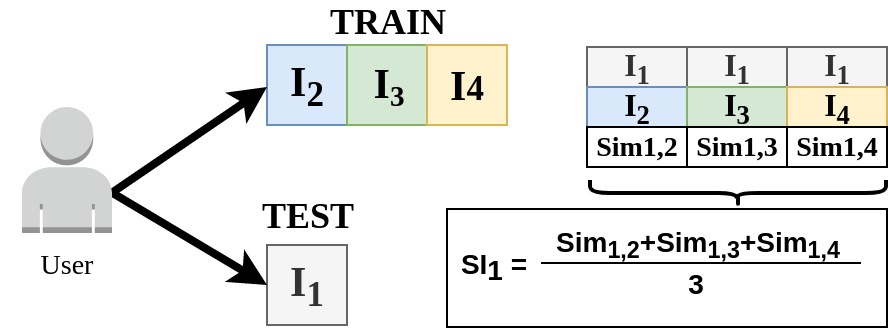}
\caption{\label{fig:onto_example}Example of ONTO algorithm. I\textsubscript{1} is a test item, I\textsubscript{2}, I\textsubscript{3} and I\textsubscript{4} are train items. The semantic similarity is calculated for each pair of test-train items. The score for I\textsubscript{1} (SI\textsubscript{1}) is the mean of the similarities of each test-train pair.}
\end{wrapfigure}

The CB module (ONTO algorithm) is based on ChEBI ontology. This module assigns a score S to each item in the test set, calculating the semantic similarity between each item in the train and the test sets, as shown in Figure \ref{fig:onto_example}. For calculating the similarity, we used DiShIn\footnote{\url{https://github.com/lasigeBioTM/DiShIn}} \cite{couto2019semantic}, a tool for calculating semantic similarities between the entities represented by an ontology. Semantic similarity allows measuring how close two entities are in a semantic base. When using ontologies, the semantic similarity may be measured, for example, by calculating the shortest path connecting the nodes of two entities. DiShin allows to calculate three similarity metrics: Resnik~\cite{resnik1995using}, Lin~\cite{lin1998information}, and Jiang and Conrath~\cite{jiang1997semantic}. For this work, we used the Lin metric. We intend to test the other metrics in the future.

 Whereas the CF module uses all the ratings from the train set to train the model, CB module only takes into account the ratings of each user. Using DiShin, we calculate the value of the similarity between each item in the train test and the items in test set. 

Lets I\textsubscript{1} be the item in test, and I\textsubscript{2}, I\textsubscript{3} ... I\textsubscript{n} the items in the train, with size \textit{m}, for a user U. The score S for I\textsubscript{1} (S\textsubscript{I}\textsubscript{1}) is calculated according to the Equation \ref{eq:score}. ONTO algorithm does not use any real rating of the test items when calculating the score for each item in the test set, thus we do not have the problem of introducing bias in the results.
\begin{equation}
\label{eq:score}
    S\textsubscript{I}\textsubscript{1} = \frac{Sim_{1,2} + Sim_{1,3} + ... + Sim_{1,n}}{m}
\end{equation}
For obtaining a final score (FS) for each item in the test, we combine the scores from CF module (S\textsubscript{CF}) and CB module (S\textsubscript{CB}), into a Hybrid recommendation approach, according to Equation \ref{eq:fs}. Our goal is to prove that by combining both modules, we can improve the results of each module separately. 
\begin{equation}
\label{eq:fs}
    FS\textsubscript{I}\textsubscript{1} = S\textsubscript{CF}\times S\textsubscript{CB}
\end{equation}
\section{Experiments and Results}
\label{Sec:resanddis}
\subsubsection{Experiments} The data used in this work is a subset of a dataset of Chemical Compounds, CheRM, with the format of \textless user,item,rating\textgreater~\cite{barros2019using}. The users are authors from research articles, the items are Chemical Compounds present in ChEBI, and the ratings (implicit) are the number of articles the author wrote about the item\footnote{\url{https://github.com/lasigeBioTM/CheRM}}.  
The subset has 102 Chemical Compounds, 1184 authors, 5401 ratings, and a sparsity level of 95.5\%. We used a subset of CheRM because it has more than 22,000 items and there is a bottleneck in the calculation of the similarity between all the items in real time.

 The algorithms tested were ALS, BPR, ONTO, and the hybrids ALS\_ONTO and BPR\_ONTO. For ALS and BPR we tested different latent factors, achieving the best results for this data with 150 factors. We used offline methods~\cite{shani2011evaluating} for evaluating the performance of the algorithms for the top@k, with k varying between 0 and 20, with steps of 1. From the vast range of metrics for evaluating recommender algorithms, we selected Classification Accuracy Metrics (CAMet) and Rank Accuracy Metrics (RAMet). CAMet measure the relevant and irrelevant items recommended in a ranked list. Examples of CAMet are Precision, Recall, and F-Measure. RAMet measure the ability of an algorithm for recommending the items in the correct order. Some well-known RAMet are Mean Reciprocal Rank (MRR), Normalized Discount Cumulative Gain (nDCG), and Limited Area Under the Curve (lAUC), a variation of AUC~\cite{schroder2011setting}. All the selected metrics range between 0 and 1, and values closest to 1 are better. For the segmentation of the dataset, we used a cross-validation approach, by splitting users and items in 5 folds. Each iteration had 1/5 of the users and the items as test and 4/5 as train data. All the positive ratings in the test set are considered as relevant items. We considered the unrated items as negative ratings, i.e., not relevant for the users.
\begin{figure}[ht]
\includegraphics[width=1\textwidth]{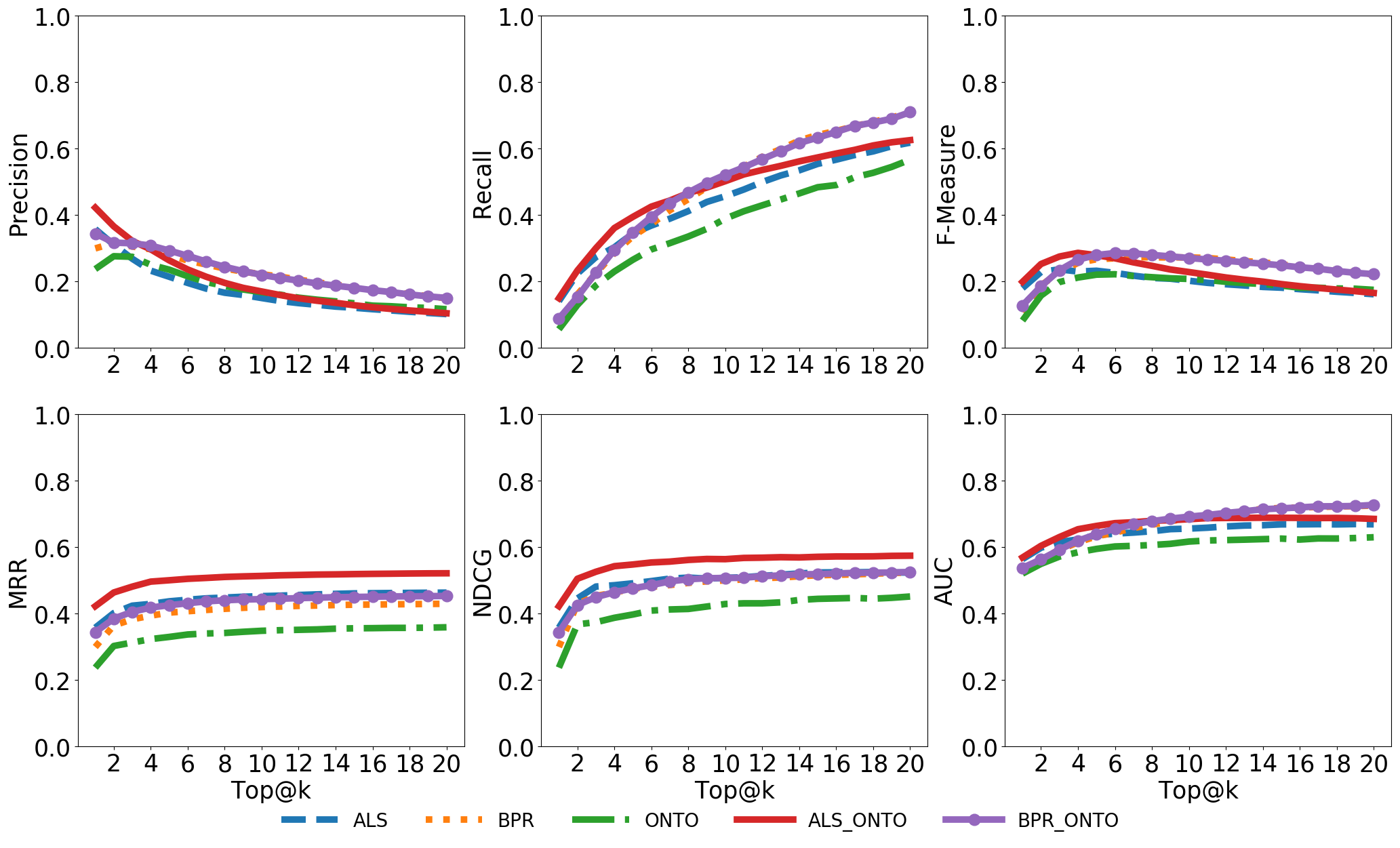}
\caption{Results comparing ALS, BPR, ONTO and the hybrids ALS\_ONTO and BPR\_ONTO, for Precision, Recall, F-measure, MRR, nDCG, and lAUC.}
\label{fig:plots}
\end{figure}

\subsubsection{Results} 
We present the results of this study in Figure \ref{fig:plots}, for all the algorithms and all the metrics described previously. Analysing Figure \ref{fig:plots}, the ONTO algorithm alone has the lowest results in all metrics. Nevertheless, in metrics such as Precision, Recall and F-measure, it follows the trend of the other algorithms, and when measuring these metric for the top@20, the results are similar. ONTO has the advantage of being a CB algorithm, therefore it does not have the problem of cold start for new items. ALS and BPR cannot be used if the item in the test set is not in the train set at least once (at least one author in the train set wrote about this Chemical Compound). 

Between ALS and BPR, ALS achieved the best results. Since BPR is an algorithm for ranking, it was expected to obtain better results. We believe this is due the fact that the dataset has a large number of ratings equal to one, and many items have the same relevance (difficult to rank).

The approach with the best results in most of the metrics is the Hybrid ALS\_ONTO. The use of ALS and ONTO algorithms together has a particularly positive effect on the metrics measuring the ranking accuracy (MRR, nDCG and AUC), especially for MRR, with an increase of 6.7\% when comparing the ALS algorithm and the Hybrid ALS\_ONTO. This means that ONTO reorder ALS scores in a way that the first results in the top@k are more relevant. 

These are preliminary results. The study needs to be replicated with the full CheRM dataset, and we need to perform more studies to see the real impact for the cold start problem. Nevertheless, the results seem promising, in the one hand for improving the relevant recommendations provided (CAMet), and on the other hand in enhancing the position of the most relevant items in a ranked list (RAMet). Our Hybrid algorithm may be applied to other areas, for example, for genes, phenotypes, and diseases, provided that exists an ontology for these items.   

%\newpage
\section{Conclusion}
\label{Sec:conc}
In this work, we presented a Hybrid recommendation model for recommending Chemical Compounds, based on CF algorithms for implicit data and a CB algorithm based on semantic similarity of the Chemical Compounds using the ChEBI ontology. 
The obtained results support our hypothesis that by using the semantic similarity between the Chemical Compounds, the results of state-of-the-art CF algorithms can be improved. For future work we intend to increase the length of the dataset, to test other similarity metrics, and to test other alternatives to calculate the final score of the Hybrid algorithm.

\noindent

%
% ---- Bibliography ----
%
% BibTeX users should specify bibliography style 'splncs04'.
% References will then be sorted and formatted in the correct style.
%
\bibliographystyle{splncs04}
\bibliography{bibliography}

%
%\begin{thebibliography}{8}
%\bibitem{ref_article1}
%Author, F.: Article title. Journal \textbf{2}(5), 99--110 (2016)

%\bibitem{ref_lncs1}
%Author, F., Author, S.: Title of a proceedings paper. In: Editor,
%F., Editor, S. (eds.) CONFERENCE 2016, LNCS, vol. 9999, pp. 1--13.
%Springer, Heidelberg (2016). \doi{10.10007/1234567890}

%\bibitem{ref_book1}
%Author, F., Author, S., Author, T.: Book title. 2nd edn. Publisher,
%Location (1999)

%\bibitem{ref_proc1}
%Author, A.-B.: Contribution title. In: 9th International Proceedings
%on Proceedings, pp. 1--2. Publisher, Location (2010)

%\bibitem{ref_url1}
%LNCS Homepage, \url{http://www.springer.com/lncs}. Last accessed 4
%Oct 2017
%\end{thebibliography}
\end{document}